\documentclass[sigconf,9pt]{acmart}

\usepackage{amsmath,amssymb,amsfonts}
\usepackage[linesnumbered,ruled,vlined]{algorithm2e}
\usepackage{graphicx}
\usepackage{textcomp}
\usepackage{xcolor}
\usepackage{comment}
\usepackage{stackengine}
\usepackage{pifont}
\usepackage{color}
\usepackage{svg}
\usepackage{tabularray}
\usepackage{hyperref}
\usepackage{subcaption}
\hypersetup{hypertex=true,
colorlinks=true,
linkcolor=black,
anchorcolor=black,
citecolor=black}

\AtBeginDocument{%
  }

\copyrightyear{2026}
\acmYear{2026}
\setcopyright{cc}
\setcctype{by}
\acmConference[DAC '26]{63rd ACM/IEEE Design Automation Conference}{July 26--29, 2026}{Long Beach, CA, USA}
\acmBooktitle{63rd ACM/IEEE Design Automation Conference (DAC '26), July 26--29, 2026, Long Beach, CA, USA}
\acmDOI{10.1145/3770743.3804144}
\acmISBN{979-8-4007-2254-7/2026/07}

\begin{document}

\title{eLLM: Elastic Memory Management Framework for Efficient LLM Serving
}
\vspace{-0.2cm}

\author{Jiale Xu}
\authornote{Equal contribution}
\affiliation{%
\vspace{-1pt}  
  \institution{Shanghai Jiao Tong University \\   Shanghai Qi Zhi Institute}
  \country{China}
}

\author{Yi Xiong}
\authornotemark[1]
\affiliation{%
\vspace{-1pt}  
  \institution{University of Science and Technology of China}
  \country{China}
}

\author{Rui Zhang}
\authornotemark[1]
\affiliation{%
\vspace{-1pt}  
  \institution{Unaffiliated}
  \country{China}
}
\vspace{-3pt}  
\author{Cong Guo}
\authornote{Corresponding author}
\affiliation{%
\vspace{-1pt}  
  \institution{Shanghai Jiao Tong University \\  Shanghai Qi Zhi Institute}
  \country{China}
}
\vspace{-3pt}  
\author{Zihan Liu}
\affiliation{%
\vspace{-1pt}  
  \institution{Shanghai Jiao Tong University \\  Shanghai Qi Zhi Institute}
  \country{China}
}
\vspace{-3pt}  
\author{Yangjie Zhou}
\affiliation{%
\vspace{-1pt}  
  \institution{National University of Singapore}
  \country{China}
}
\vspace{-3pt}  
\author{Weiming Hu}
\affiliation{%
\vspace{-1pt}  
  \institution{Shanghai Jiao Tong University \\  Shanghai Qi Zhi Institute}
  \country{China}
}
\vspace{-3pt}  
\author{Hao Wu}
\affiliation{%
\vspace{-1pt}  
  \institution{Unaffiliated}
  \country{China}
}
\vspace{-3pt}  
\author{Boyu Li}
\affiliation{%
\vspace{-1pt}  
  \institution{University of Science and Technology of China}
  \country{China}
}
\vspace{-3pt}  
\author{Junping Zhao}
\authornotemark[2]
\affiliation{%
\vspace{-1pt}  
  \institution{Unaffiliated}
  \country{China}
}
\vspace{-3pt}  
\author{Minyi Guo}
\affiliation{%
\vspace{-1pt}  
  \institution{Shanghai Jiao Tong University \\ Shanghai Qi Zhi Institute}
  \country{China}
}
\vspace{-3pt}  
\author{Zongwei Zhu}
\authornotemark[2]
\affiliation{%
    \vspace{-1pt}  
  \institution{University of Science and Technology of China}
  \country{China}
}
\vspace{-3pt}  
\author{Xuehai Zhou}
\affiliation{%
\vspace{-1pt}  
  \institution{University of Science and Technology of China}
  \country{China}
}
\vspace{-3pt}  
\author{Jingwen Leng}
\affiliation{%
\vspace{-1pt}  
  \institution{Shanghai Jiao Tong University \\  Shanghai Qi Zhi Institute}
  \country{China}
}







\renewcommand{\shortauthors}{Xu{*}, Xiong{*}, Zhang{*}, Guo et al.}

\begin{abstract}
GPU memory management is critical for efficient Large Language Model (LLM) serving. LLM memory usage primarily comprises weights, activations, and KV caches. While weights are static, activations and KV caches exhibit dynamic and unpredictable behavior, posing significant memory management challenges. Modern LLM serving systems address this through a dual-level approach: activations inherit static tensor abstractions from deep learning frameworks, while KV caches employ specialized page-table virtualization (i.e., PagedAttention). Although this reduces KV cache fragmentation, the fundamental isolation between activation and KV cache management prevents memory sharing across these spaces, leading to suboptimal utilization and 20\% throughput degradation.

To address these limitations, we propose eLLM, an elastic memory management framework. The core components of eLLM include:(1) Virtual Tensor Abstraction: Decouples the virtual address space of tensors from physical GPU memory, creating a unified and flexible memory pool;(2) Elastic Memory Mechanism: Dynamically adjusts memory allocation through runtime memory inflation and deflation, and leverages CPU memory as an extensible buffer;(3) Lightweight Scheduling Strategy: Employs Service-Level Objective (SLO)-aware policies to optimize memory utilization and effectively balance performance trade-offs under stringent SLO constraints.
Comprehensive evaluations demonstrate that eLLM outperforms state-of-the-art systems, achieving up to 2.32$\times$ higher throughput.
\vspace{-0.3cm}

\end{abstract}
\vspace{-0.2cm}
\vspace{-0.5cm}

\maketitle

\vspace{-0.2cm}
\section{Introduction}\label{sec:introduction}
Large Language Models (LLMs)~\cite{DBLP:llama} process billions of requests daily in modern AI services. With model sizes growing exponentially while GPU memory capacity increases only incrementally, GPU memory has emerged as the critical bottleneck for high-performance LLM serving.
First, the memory-bound nature of auto-regressive generation necessitates request batching to improve arithmetic intensity and amortize framework overhead. However, limited GPU memory directly constrains batch sizes, capping system throughput~\cite{DBLP:vllm}. Second, when GPU memory is insufficient to accommodate the incoming requests, these requests must queue until running requests complete and release memory, which significantly degrades response latency~\cite{layerkv}.


\begin{figure*}[t]
    \vspace{-0.3cm}
    \centering
    \includegraphics[width=0.6\textwidth]{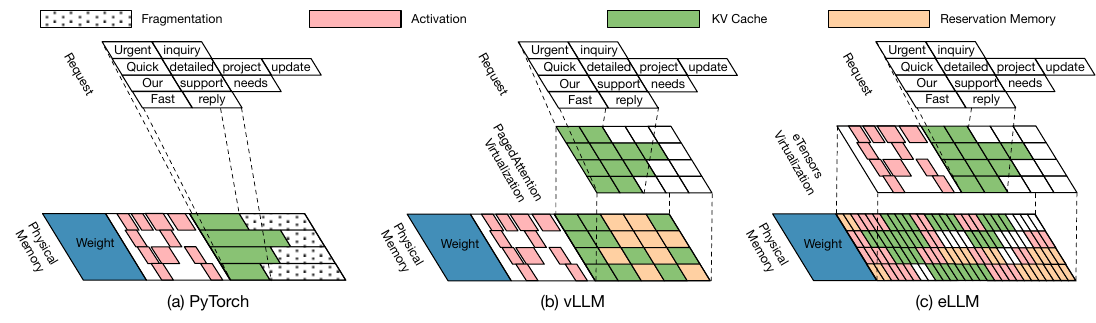} 
    \vspace{-0.4cm}
    \caption{(a) Early LLM systems~\cite{DBLP:orca} inherit the static-tensor abstraction from deep learning frameworks (e.g., Pytorch~\cite{DBLP:torch}), failing to handle runtime KV cache tensor expansion, causing fragmentation. (b) By virtualizing the KV cache, vLLM~\cite{DBLP:vllm} alleviates memory fragmentation but separates activations and the KV cache into different abstraction levels, thereby isolating activation from the KV cache space.
    (c) eLLM independently manages the KV cache and activations at the logical level while consolidating memory into a unified physical pool, thereby maximizing memory utilization.}
    \label{fig:isolation} 
    \vspace{-0.4cm}
\end{figure*}

\textbf{Status Quo.} The core memory components of LLM inference include model weights, activations, and KV caches. Among these, activations and KV caches exhibit highly \textbf{dynamic} and \textbf{unpredictable} behavior. Their dynamic nature stems from frequent allocation and de-allocation during inference, while their unpredictability arises from tensor sizes that vary with the uncertain lengths of input prompts and generated sequences.

To manage these tensors, early LLM serving systems~\cite{DBLP:orca} rely on static tensor abstractions inherited from deep learning frameworks (e.g., PyTorch~\cite{DBLP:torch}). As this abstraction assumes that tensor sizes remain constant throughout their lifecycles, these systems naturally allocate contiguous physical memory blocks and employ memory-pool-based reuse strategies for efficient memory management.
However, persisting with this abstraction for KV caches, which continuously expand and have unpredictable sizes, inevitably results in severe memory fragmentation, as shown in Figure~\ref{fig:isolation}(a).

To solve this, vLLM~\cite{DBLP:vllm} introduces PagedAttention, a virtual memory abstraction tailored for KV caches. 
During system initialization, it pre-allocates GPU memory into two pools: a reserved portion for model weights and activations (with activations sized according to maximum sequence length), and the remaining memory organized as fixed-size KV blocks that are dynamically allocated to requests on demand. This strategy significantly reduces fragmentation, as shown in Figure~\ref{fig:isolation}(b), making PagedAttention the de facto standard for modern LLM serving systems.

\textbf{Limitations.} However, PagedAttention creates isolation between activations and KV cache spaces. Activations rely on the static tensor abstraction of framework level with direct physical memory allocation, while KV caches are managed through page-table virtualization built atop these abstractions. 
Due to this abstraction hierarchy difference, the two memory spaces cannot share resources, even when the KV cache space is exhausted while the activation memory sits idle (or vice versa), the underutilized space cannot be repurposed.
We refer to this phenomenon as \emph{\textbf{space-wise internal fragmentation}}. 

\begin{figure}[t]
    \centering
    \begin{subfigure}{0.40\textwidth}
        \centering
        \includegraphics[width=0.8\textwidth]{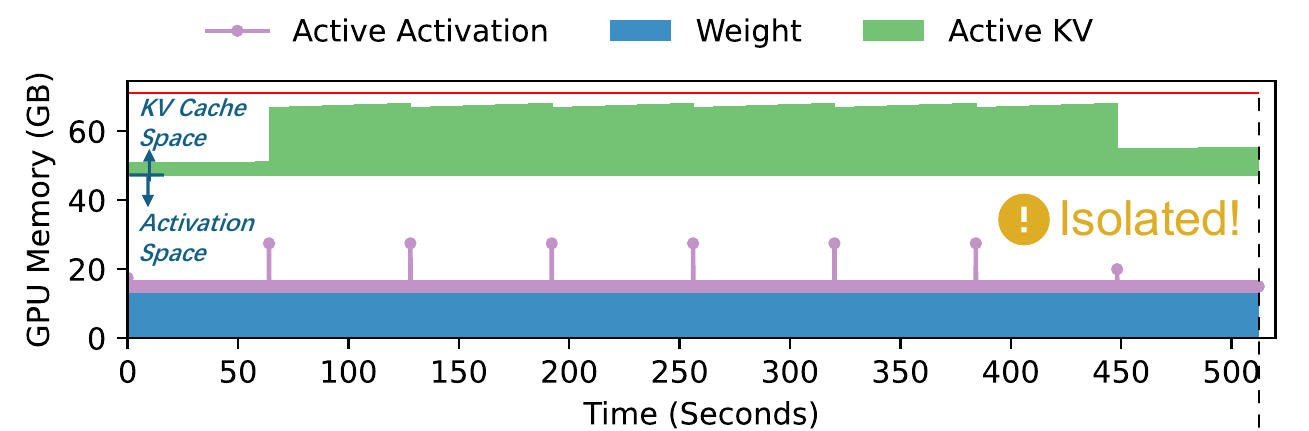}
        \vspace{-0.25cm}
        \caption{vLLM (KV cache-centric) memory snapshot.}
        \label{fig:intro2}
    \end{subfigure}
    \hfill
    \begin{subfigure}{0.40\textwidth}
        \centering
        \includegraphics[width=0.8\textwidth]{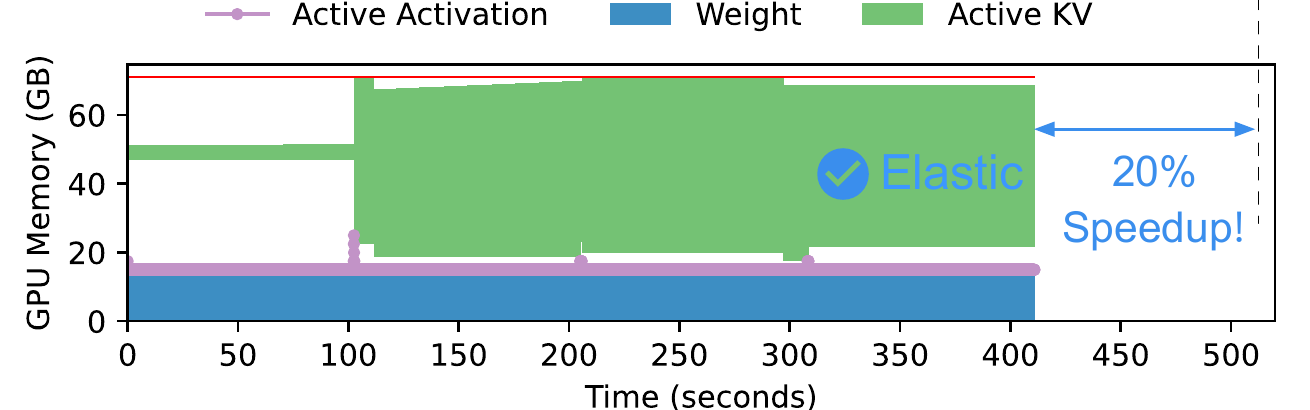}
        \vspace{-0.25cm}
        \caption{eLLM (ours) memory snapshot.}
        \label{fig:intro3}
    \end{subfigure}
    \vspace{-0.3cm}
    \caption{(a) vLLM’s isolated allocation for activations and KV cache in separate spaces causes underutilization and suboptimal performance; (b) \texttt{eLLM} enables elastic memory allocation, maximizing utilization and achieving a 1.2$\times$ speedup.}
    \vspace{-0.5cm}
    \label{fig:intro} 
\end{figure}

The phenomenon has become increasingly prominent throughout the entire inference phase.
\textbf{Prefill phase:}
As model context lengths surge from thousands~\cite{DBLP:llama} to millions~\cite{DBLP:LM-Infinite,DBLP:LWM}, the pre-reserved activation space grows proportionally. For example, increasing the maximum context of MTP-30B from 2K to 200K on the same A100 GPUs increases the activation reservation from only 0.3\% to 30.8\% of GPU memory. However, real-world prompt lengths follow a typical long-tail distribution; for instance, 99\% of prompts fall within 2K tokens in ShareGPT~\cite{Sharegpt}. This leads to significant space-wise internal fragmentation, as more than \textbf{90\%} of prefill batches use less than \textbf{30\%} of the model’s maximum context length.
\textbf{Decoding phase:}
Unlike the prefill phase, which processes a large number of tokens at once, each decoding step handles only a small number of tokens, resulting in even lower activation space utilization ($\sim$1\%). As advanced inference techniques such as Chain-of-Thought (CoT)~\cite{DBLP:cot} become widely adopted, decoding cycles become longer, and this extremely low utilization increasingly dominates the inference process.

Advanced LLM serving systems may mitigate this fragmentation from the scheduling layer, but their limitations have also become widely recognized.
\textbf{Chunked Prefill (CP)}~\cite{DBLP:Sarathi-Serve} reduces peak activation memory by partitioning requests into fixed-size chunks, thereby reducing pre-reserved activation space.  However, it suffers from an inherent drawback refered as \emph{read amplification~\cite{DBLP:Mnemosyne}}: Chunked prefill processing incurs redundant KV cache reloading, as each new chunk requires accessing all previously generated KV caches. This memory overhead scales super-linearly with sequence length. 
\textbf{Prefill–decode disaggregation systems}~\cite{DBLP:Mooncake,DBLP:split-wise,DBLP:DistServe} separate inference phases across different GPUs, reducing activation memory reservation on decode instances. However, they also exhibit four limitations~\cite{DBLP:journals/corr/abs-2504-19867}:
(1) Model Replication Overhead: Full model weights must be replicated across prefill and decode instances, imposing substantial memory redundancy in small-scale deployments;
(2) Memory Utilization Skew: Decode instances experience memory saturation from long-lived KV caches, while prefill instances remain significantly underutilized;
(3) Inter-instance Communication Cost: KV cache transfer from prefill to decode instances incurs non-trivial network latency and bandwidth consumption;
Unlike these approaches, we directly address the space-wise internal fragmentation at the memory management level, avoiding their limitations and remaining orthogonal to them.

\textbf{Our solution.} 
We propose \textbf{\texttt{eLLM}}, an elastic memory management framework. As Figure~\ref{fig:isolation}(c) shows, at the logical level, eLLM introduces unified virtual tensors that treat KV caches and activation tensors as independent logical entities at the same abstraction level, enabling tailored memory management strategies for their distinct access patterns. At the physical level, eLLM implements elastic memory mechanisms through inflation and deflation operations to dynamically adjust memory allocation. 
As illustrated in Figure~\ref{fig:intro}, activations and KV caches can mutually "borrow" memory from each other, achieving approximately 20\% throughput improvement and demonstrating the effectiveness of our elastic memory management. Additionally, eLLM leverages CPU memory as an extensible buffer for GPU memory, enhancing memory flexibility and alleviating GPU memory pressure. Finally, eLLM incorporates a lightweight scheduling strategy that coordinates elastic memory management to optimize memory allocation, task scheduling, and resource transitions. Extensive evaluation across diverse models and workloads demonstrates that eLLM achieves significant performance improvements, with inference throughput gains up to 2.32$\times$.

In summary, we make the following contributions:
\begin{itemize}
    \item We identify the space-wise internal fragmentation problem in existing LLM serving systems.
    \item We propose eTensor, a virtual tensor abstraction that enables memory transfers between activations and KV cache.
    \item We design elastic memory mechanisms: inflation/deflation for GPU-internal rebalancing and SLO-aware CPU buffering.
    \item Comprehensive evaluations demonstrate that eLLM outperforms state-of-the-art LLM serving systems.
\end{itemize}
\vspace{-0.3cm}

\section{Design}
eLLM comprises three core components organized in a bottom-up architecture: Virtual Tensor Abstraction, Elastic Memory Mechanism, and Lightweight Scheduling Strategy.
\ding{182} eLLM introduces a unified virtual tensor abstraction for both KV caches and activations, decoupling them from physical resources while preserving their distinct access patterns.
\ding{183} Building on this abstraction, eLLM implements an Elastic Memory Mechanism enabling: (1) intra-GPU elasticity through inflation/deflation operations between KV caches and activations, and (2) GPU-CPU elasticity using CPU memory as an overflow buffer.
\ding{184} eLLM employs a Lightweight Scheduling Strategy that coordinates these elastic mechanisms with SLO-aware adaptive buffer sizing.

\vspace{-0.3cm}
\subsection{Virtual Tensor Abstraction}
\label{sec:vta}
\begin{figure}[t]  
    \centering  
    \includegraphics[width=0.6\linewidth]{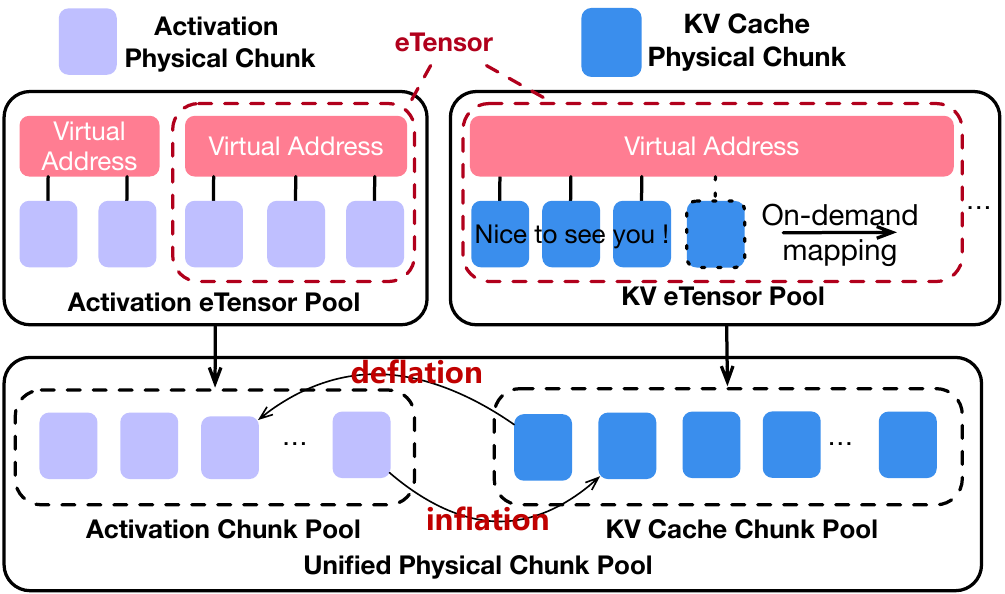} 
    \caption{eTensor abstraction for KV cache and activations. }
    \label{fig:etensor}  
    \vspace{-0.2cm}
\end{figure}

The root cause of space-wise internal fragmentation lies in the inconsistent abstraction levels between activations and KV caches. 
Two naive approaches could potentially unify these abstractions: (1) Fine-grained KV cache management delegated to the framework would introduce substantial metadata overhead and compromise the sharing structure of KV caches; (2) A type-agnostic unified virtualization abstraction would ignore the distinct characteristics of activations and KV caches, precluding workload-specific optimizations.
Based on this observation, eLLM introduces eTensor, a novel tensor abstraction with two specialized types, KV eTensor and activation eTensor. Each implements tailored mechanisms for its specific access patterns while sharing a unified memory pool for inter-conversion, as illustrated in Figure~\ref{fig:etensor}.

\vspace{-0.1cm}
\subsubsection{Virtualization Design}
KV caches are inherently large, regular memory blocks with predictable growth patterns, low access frequency, and persistent retention throughout inference. Therefore, KV eTensor pre-allocates virtual address space equal to the maximum context length for each request, ensuring logical continuity that facilitates sequential access and index computation, while physical blocks are allocated on-demand during actual writes to avoid unnecessary memory consumption.
In contrast, activation tensors consist of smaller memory blocks with short lifespans and high access frequency. Activation eTensor employs variable-sized virtual address segments requiring frequent fine-grained management. Both eTensor types align their virtual address segments (termed tensor slots) strictly to physical block granularity, achieving better balance between access efficiency and fragmentation control.


\vspace{-0.1cm}
\subsubsection{Memory Management Strategy}
To mitigate the potential overhead introduced by tensor virtualization via virtual-to-physical address mappings, eLLM implements tailored memory-pool management strategies for each eTensor type. 
Rather than immediately unmapping eTensor instances from physical resources at the end of their lifecycles, the system marks them as mapped and available tensor slots, and tracks their mapping sizes, thereby enabling efficient memory reuse.
The KV eTensor pool employs a \textit{\textbf{Best-Fit algorithm}} for incoming requests. Given the set of available pre-mapped tensor slots $ R=\{r \mid r \in Mapped \wedge r.state=Available \}$ and a target memory size $s$, the algorithm prioritizes selecting the smallest feasible slot satisfying the requirement $r_i=\arg \min _{r_j \in R}\left\{\operatorname{size}\left(r_j\right) \mid \operatorname{size}\left(r_j\right) \geq s\right\}$.
If no such slot exists, on-demand mapping is triggered.
The Activation eTensor pool retains the framework’s native \textit{\textbf{Best-Fit with Coalescing (BFC) strategy}}.


\vspace{-0.3cm}
\subsection{Elastic Memory Mechanism}
\label{sec:emm}

The elastic memory mechanism aims to enhance memory utilization and system performance under LLM workloads.
It introduces two levels of elasticity: intra-GPU memory \textit{inflation} and \textit{deflation}, which enable timely adjustment of physical memory allocation for activation and KV cache in response to the dynamic workloads; and GPU-CPU \textit{offloading} and \textit{fetching}, which employ CPU memory as an elastic buffer to further alleviate the memory pressure on GPU. 

\vspace{-0.1cm}
\subsubsection{Memory Inflation and Deflation}
Memory \textit{inflation} and \textit{deflation} operations are the core mechanisms of eLLM, enabling it to break memory isolation through mapping relation propagation, which dynamically maintains the binding relationship between virtual address spaces and physical memory chunks, built upon the eTensor abstraction.

Inflation operation dynamically expands the physical memory capacity of the KV cache by borrowing from the active memory pool, much like inflating a balloon.
This process involves the following steps:
\ding{182} Inflation trigger: upon KV cache allocation requests, the system first verifies whether the KV memory pool contains sufficient physical memory chunks.  If insufficient, a borrowing request is issued to the activation memory pool.  
\ding{183} Memory reclamation: the active pool fulfills the request by triggering a lightweight garbage collection (GC) phase. This GC phase identifies and unmaps physical memory chunks allocated to inactive eTensor objects.
\ding{184} Ownership transfer: reclaimed chunks are logically migrated from the activation pool to the KV cache pool through ownership transfer.
\ding{185} On-demand remapping: the GPU virtual memory manager dynamically maps these chunks to target KV eTensors' virtual address spaces.
Deflation operation is the reverse process of the above.

\begin{figure}[t]
    \centering
    \includegraphics[width=0.8\linewidth]{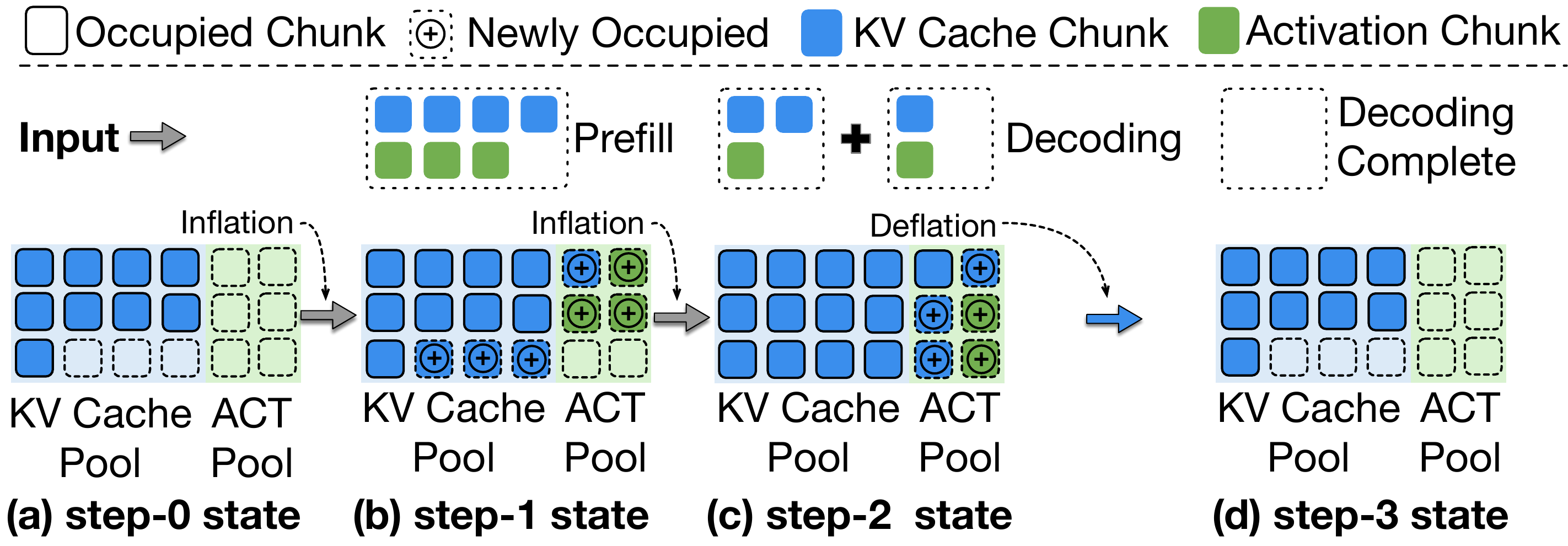} 
    \vspace{-0.5cm}
    \caption{Illustrative example of memory inflation/deflation.}
    \label{fig:example}
    \vspace{-0.4cm} 
\end{figure}

Figure~\ref{fig:example} demonstrates the intra-GPU elastic memory mechanism through an illustrative example.
(a) Initially, the GPU retains historical KV caches. In this state, existing LLM serving systems are unable to process newly arrived requests immediately due to insufficient KV cache space.
(b) With memory inflation, eLLM can handle prefill requests promptly by borrowing idle activation chunks.
(c) Through memory inflation, eLLM enables larger batch processing even under high memory pressure, thereby improving computational resource utilization given the memory-bound nature of the decoding phase.
(d) With memory deflation, eLLM can return the borrowed memory. In practice, this process is triggered lazily to avoid unnecessary overhead.

\vspace{-0.1cm}
\subsubsection{Memory Offloading and Fetching}
eLLM further incorporates GPU-CPU elastic memory management to maintain service responsiveness under memory pressure.
In online LLM serving, optimizing system response latency (i.e., TTFT) faces more challenges compared to decoding latency (i.e., TPOT): (1) more memory contention leading to severe request queuing delays, and (2) the computational complexity of the prefill phase scaling superlinearly.
Therefore, eLLM enables proactively offloading KV cache for part of the requests to CPU DRAM during the prefill phase. This lowers the memory admission barrier for request execution, effectively reducing queuing delays to improve TTFT. 
Additionally, it can aggregate larger decoding batch sizes, thereby increasing throughput.

Its technical feasibility is based on the following observations.
Firstly, newly generated KV caches that are not immediately needed can be proactively offloaded and fetched back only when their corresponding requests are scheduled for decoding.
Secondly, the multi-layer structure of Transformers naturally supports overlapping computation and communication through layer-wise pipelining.
Thirdly, migration of KV caches requires only $O(N)$ linear communication overhead, whereas the prefill stage’s self-attention computation incurs $O(N^2)$ complexity.
For example, in practical workloads with A100 GPUs and the LLaMA-3-8B model, offloading overhead can be completely hidden.
However, while CPU buffer improves request queuing, it inherently introduces a prefill-preference that adversely affects TPOT.
Therefore, eLLM introduces a simple yet effective policy to trade off between SLO metrics(§\ref{subsubsec:SLO}).

\vspace{-0.3cm}
\subsection{Lightweight Scheduling Strategy}
\label{sec:scheduler}
To address memory inefficiency in LLM serving, eLLM introduces a lightweight yet effective scheduling algorithm. 

\vspace{-0.2cm}
\subsubsection{Request Scheduling}

eLLM maximizes concurrent requests through elastic memory allocation while adhering to memory constraints. Capitalizing on the substantial variation in activation memory demands across different phases of LLM inference, we employ phase-specific allocation:
(1) Prefill phase: Requires significant activation and KV cache memory. During memory-intensive scenarios, we offload KV cache to CPU buffers, thereby reducing GPU memory pressure.
(2) Decoding phase: Given minimal activation memory requirements, we optimize resource utilization by fetching KV cache back to GPU, exploiting the reduced memory footprint.
Both phases maximize GPU memory utilization through inflation and deflation.

\vspace{-0.2cm}
\subsubsection{SLO-aware Buffer Scaling Policy}
\label{subsubsec:SLO}
CPU buffer size directly affects the trade-off between TTFT and TPOT. To achieve an optimal balance between them, it is essential to determine an appropriate buffer size. However, a fixed-size buffer is suboptimal due to LLMs' dynamic workloads. To address this, eLLM introduces the concept of a logical buffer, an abstraction of the physical buffer that dynamically adjusts the usable size within its fixed capacity. This enables flexible adjustment of buffer space without evicting stored data, thereby enhancing the system's adaptability to dynamic workloads.

Building on this, eLLM proposes an SLO-aware logical buffer scaling algorithm. If a TPOT violation is detected, the logical buffer size is reduced to limit prefill requests, thereby improving TPOT performance. Conversely, if a TTFT violation is detected, the logical buffer size is increased to improve TTFT performance. A violation event is triggered if the TTFT or TPOT exceeds the predefined SLO threshold three times within a specific scheduling iteration window (set to 5 iterations). The buffer tuning factor $\alpha$ is a hyperparameter (default $\alpha$ = 2) that controls the rate of buffer size adjustment.

\begin{figure*}[t]
    \vspace{-0.5cm}
    \centering
    \begin{subfigure}{0.21\textwidth}
        \centering
        \includegraphics[width=\textwidth]{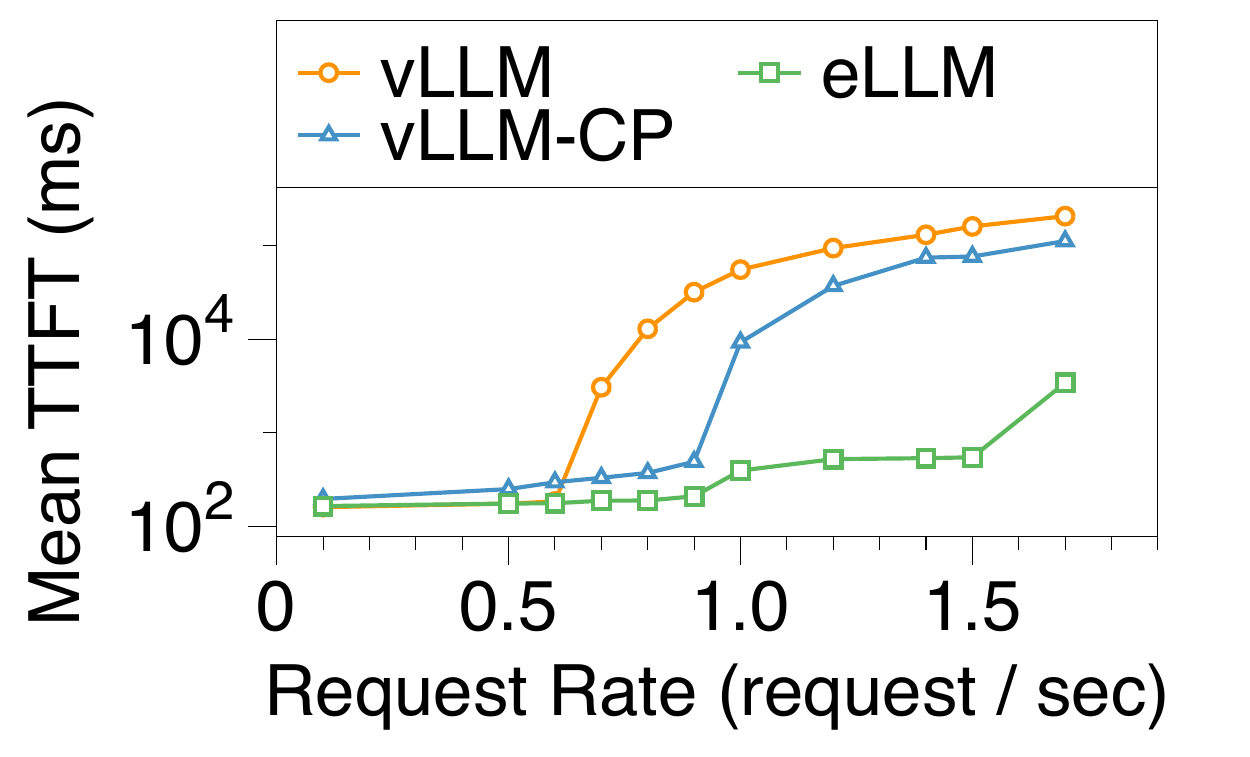}
        \captionsetup{skip=-2pt} 
        \caption{2K-2K}
        \label{fig:2k2kttft}
    \end{subfigure}
    \hfill
    \begin{subfigure}{0.21\textwidth}
        \centering
        \includegraphics[width=\textwidth]{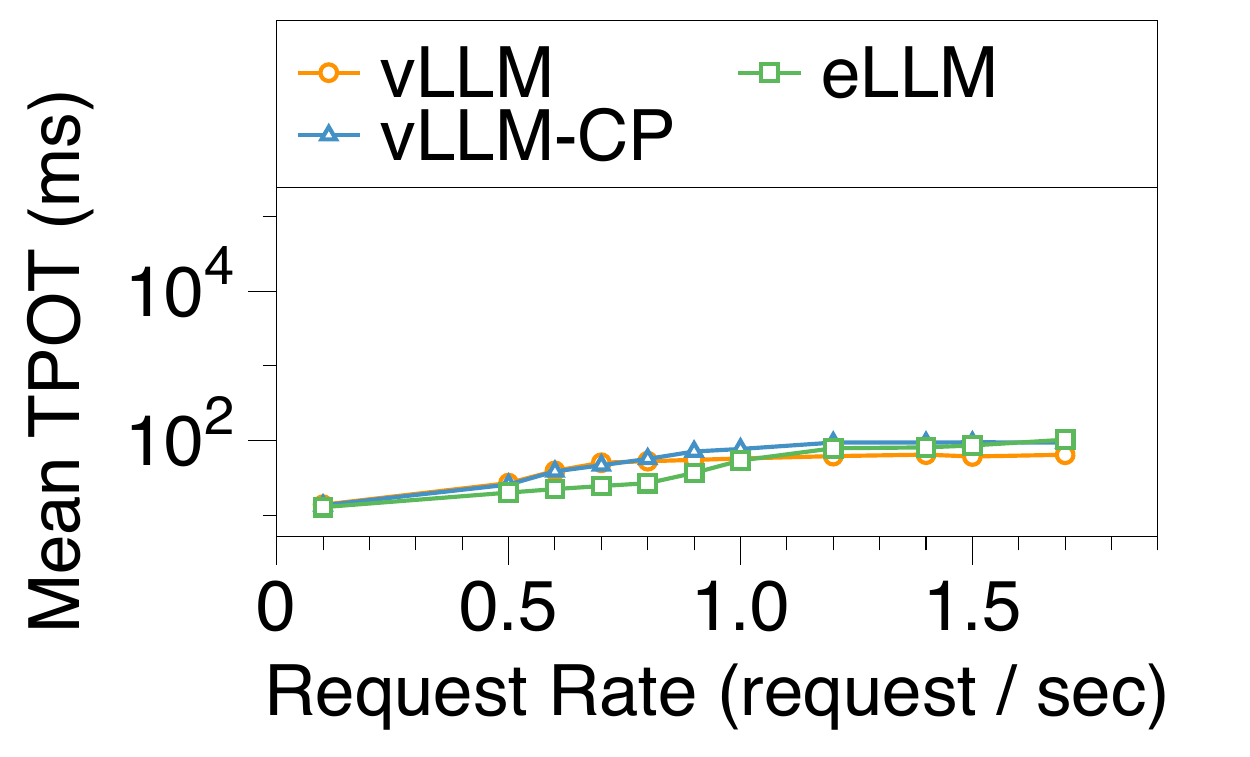}
        \captionsetup{skip=-2pt} 
        \caption{2K-2K}
        \label{fig:2k2ktpot}
    \end{subfigure}
    \hfill
    \begin{subfigure}{0.21\textwidth}
        \centering
        \includegraphics[width=\textwidth]{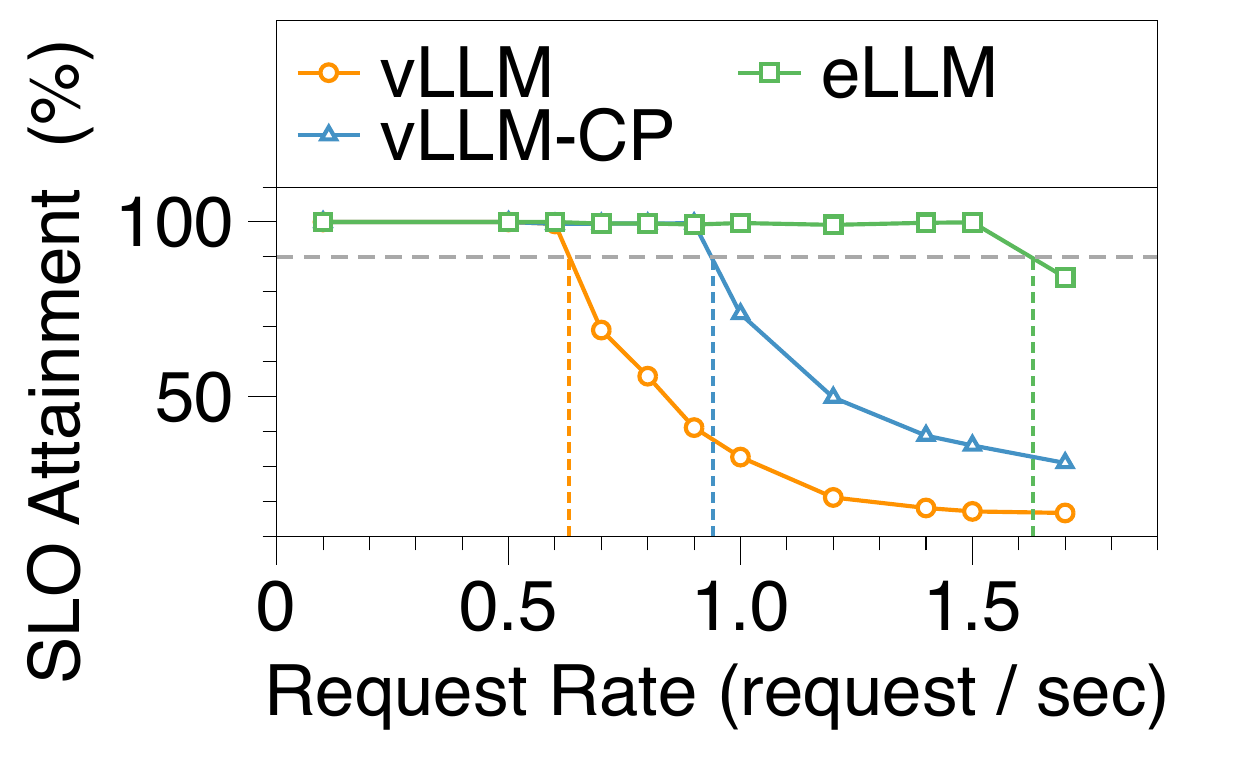}
        \captionsetup{skip=-2pt} 
        \caption{2K-2K}
        \label{fig:2k2kslo}
    \end{subfigure}
    \hfill
    \begin{subfigure}{0.21\textwidth}
        \centering
        \includegraphics[width=\textwidth]{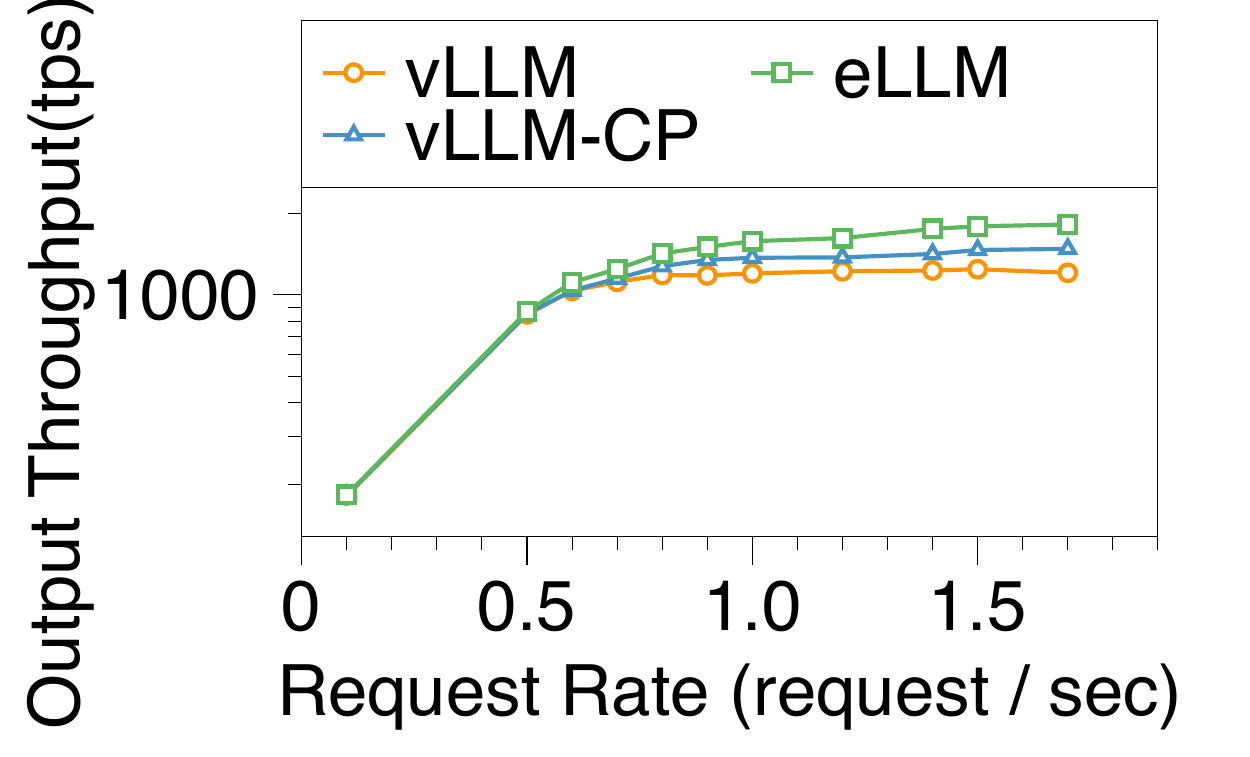}
        \captionsetup{skip=-2pt} 
        \caption{2K-2K}
        \label{fig:2k2kout}
    \end{subfigure}
    

    \begin{subfigure}{0.21\textwidth}
        \centering
        \includegraphics[width=\textwidth]{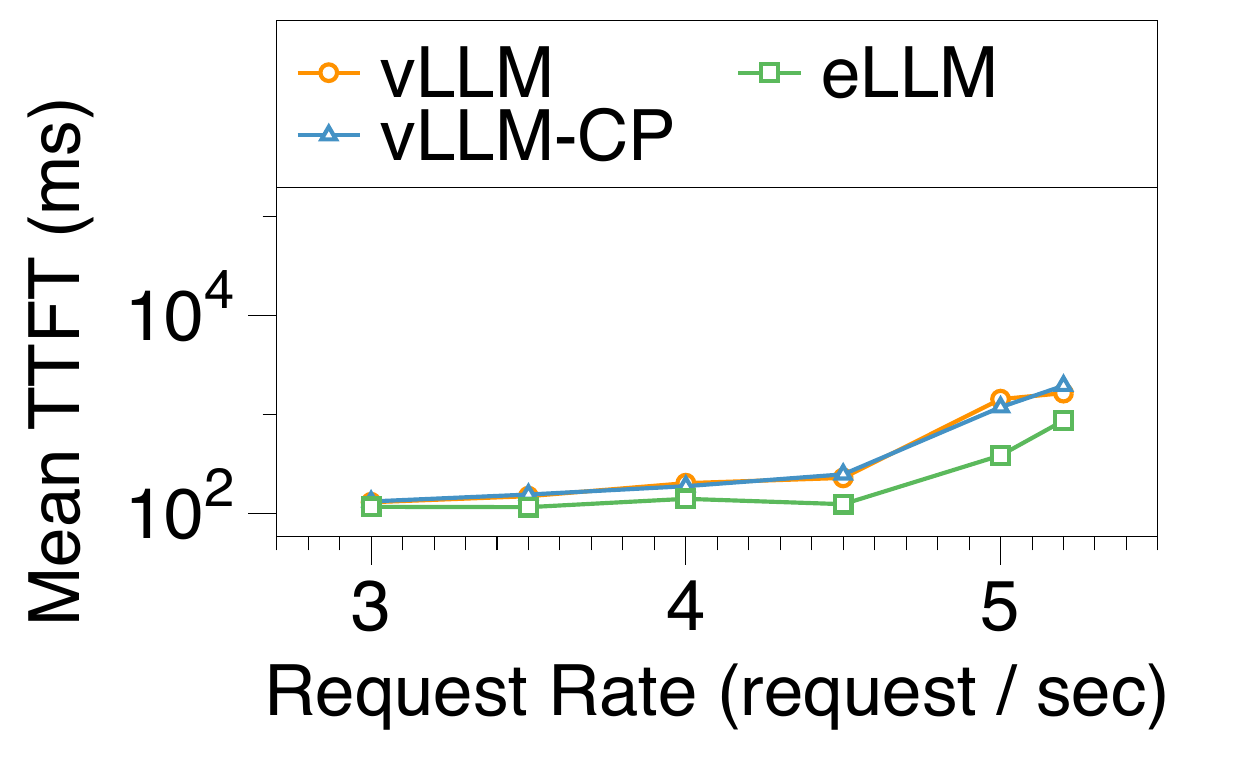}
        \captionsetup{skip=-2pt} 
        \caption{ShareGPT}
        \label{fig:shttft}
    \end{subfigure}
    \hfill
    \begin{subfigure}{0.21\textwidth}
        \centering
        \includegraphics[width=\textwidth]{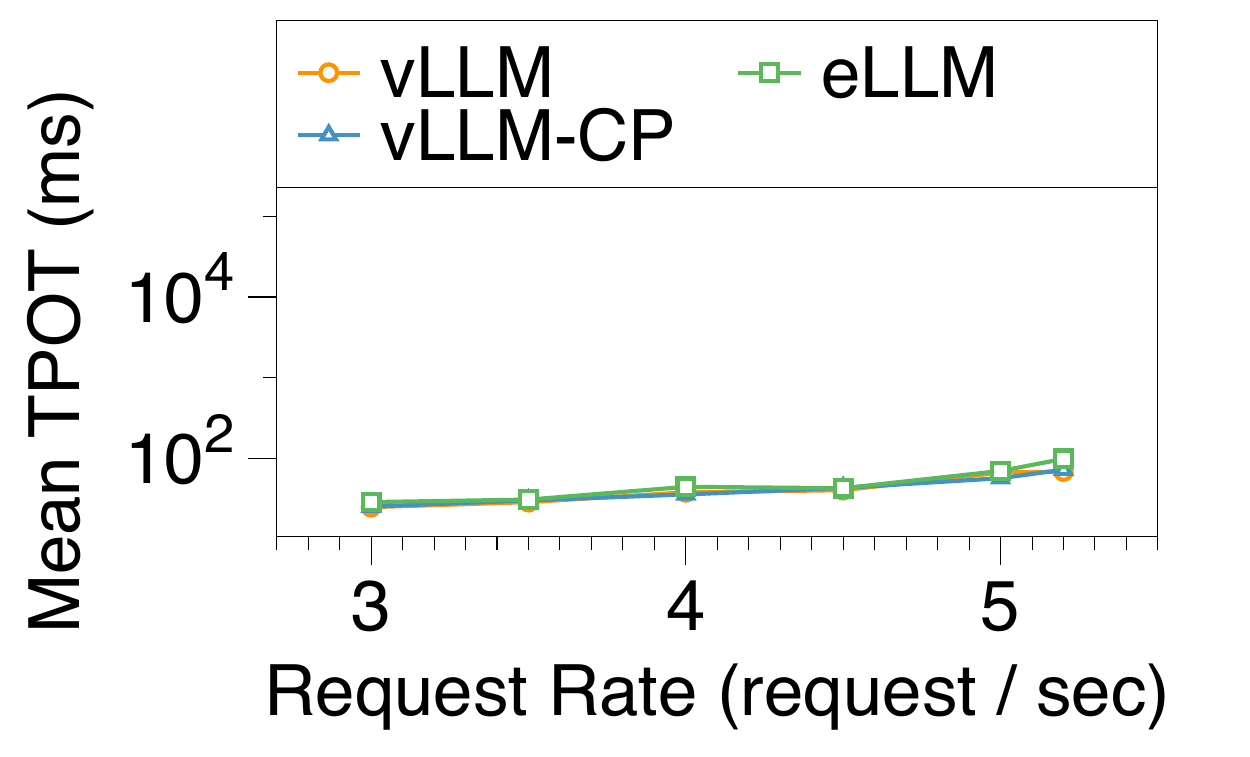}
        \captionsetup{skip=-2pt} 
        \caption{ShareGPT} 
        \label{fig:shtpot}
    \end{subfigure}
    \hfill
    \begin{subfigure}{0.21\textwidth}
        \centering
        \includegraphics[width=\textwidth]{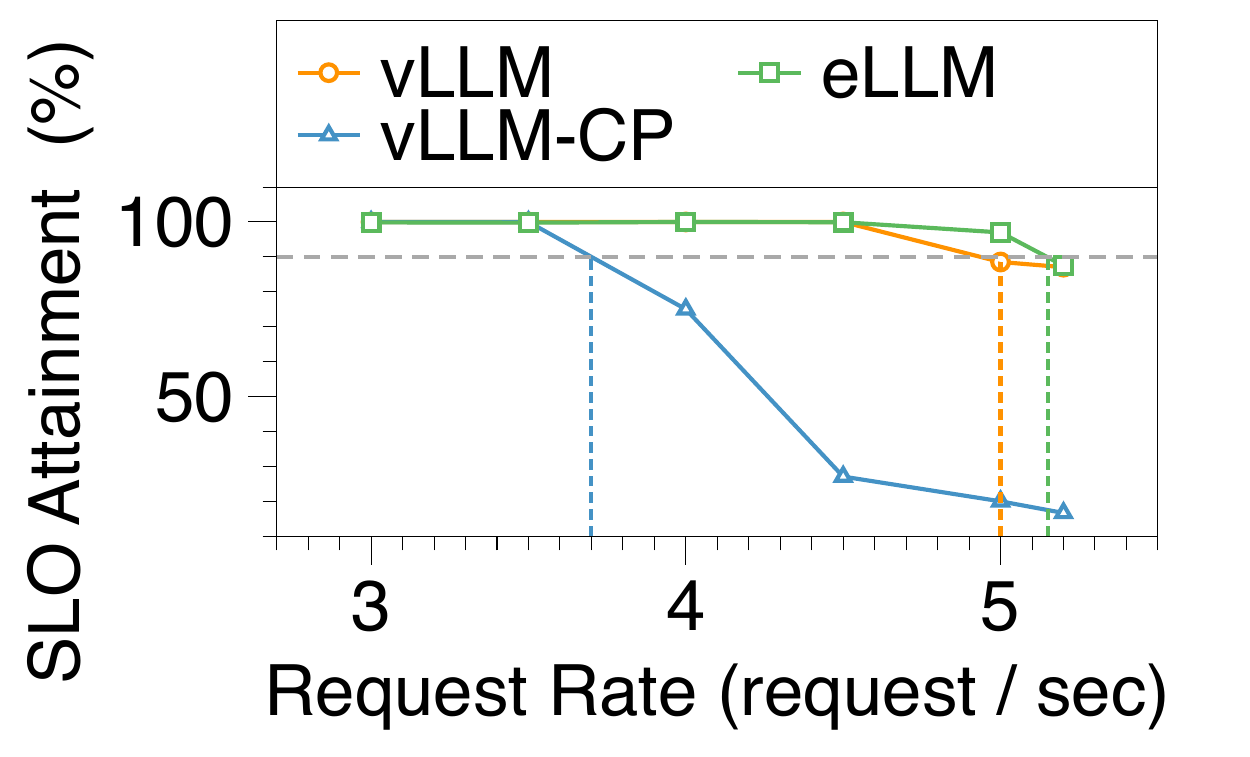}
        \captionsetup{skip=-2pt} 
        \caption{ShareGPT}
        \label{fig:shslo}
    \end{subfigure}
    \hfill
    \begin{subfigure}{0.21\textwidth}
        \centering
        \includegraphics[width=\textwidth]{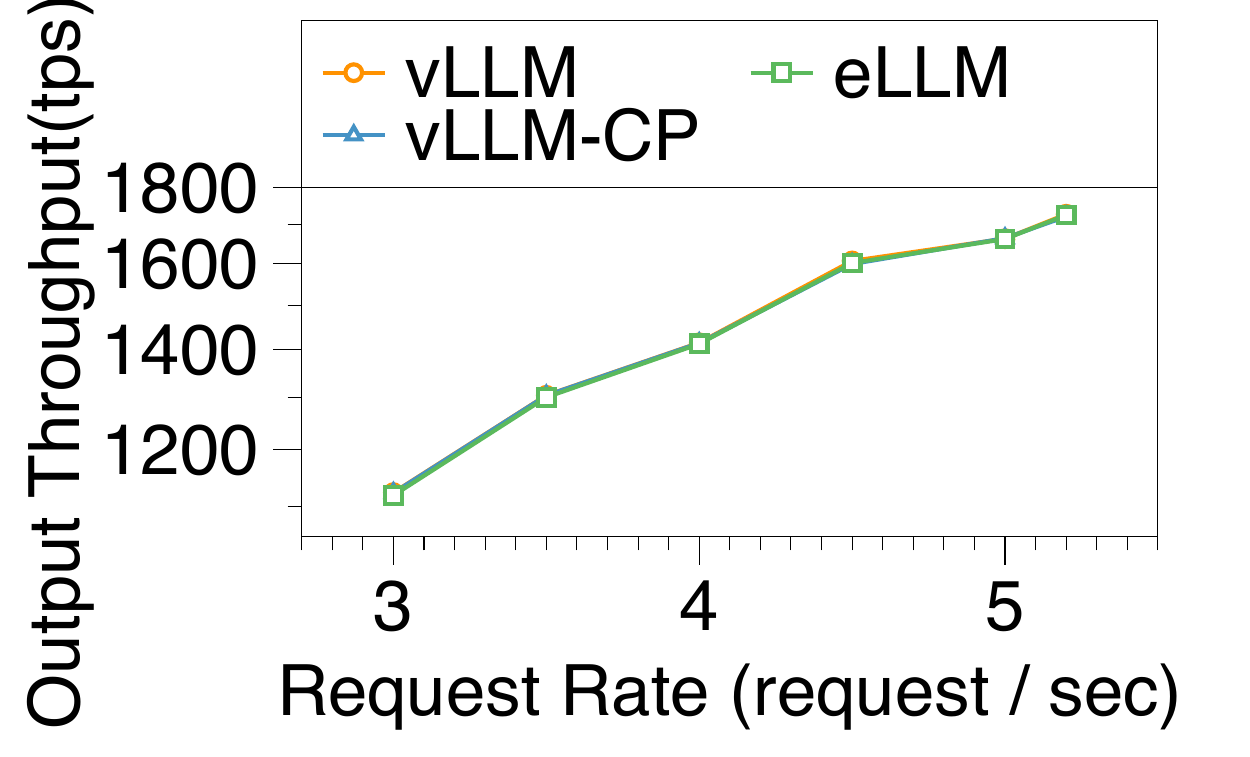}
        \captionsetup{skip=-2pt} 
        \caption{ShareGPT}
        \label{fig:shout}
    \end{subfigure}
    \vspace{-0.4cm}
    \caption{Online serving evaluation with SLO-constraints on Llama3-8B-262K model with one A100 (80GB) GPU. 
    }
    \vspace{-0.4cm}
    \label{fig:onlineserving} 
\end{figure*}

\vspace{-0.3cm}
\section{Implementation}
\label{sec:impl}
\vspace{-0.1cm}
\subsection{Overlapping (Un)mapping Overheads}
To further overlap (up)mapping overhead, we introduce several techniques, including decoding speculative pre-mapping and asynchronous unmapping.
\textbf{Decoding speculative pre-mapping.}
Due to the autoregressive nature, where each sequence generates only one token per iteration, we leverage branch speculation principles to assume each sequence will produce the next token and proactively initiate asynchronous memory allocation in advance. This approach effectively masks mapping overhead by using minimal additional memory (<50MB), thereby achieving full overlap between decoding and mapping costs.
\textbf{Asynchronous unmapping.}
When unmapping a tensor slot is triggered, the system does not require immediate unmapping of the slot followed by reassignment of its associated physical chunk. Instead, by leveraging the GPU VMM's capability to map a single physical chunk to multiple virtual addresses, the physical chunk can be initially assigned to a new tensor slot. 
The unmapping of the original slot can then be performed asynchronously, effectively overlapping the unmapping overhead.

\vspace{-0.3cm}
\subsection{eLLM System Implementation}
eLLM is built on top of vLLM v0.5.5, incorporating $\sim$4000 lines of C++ and Python code.
To support KV eTensor, we developed a C++ library and integrated it into vLLM, replacing the original KV Cache tensor object, and modified the minimum allocation granularity from a page to a physical memory chunk.
For activation eTensor management, we modify the caching allocator of PyTorch, replacing traditional memory allocation methods such as \texttt{cudaMalloc} with GPU VMM APIs.

\vspace{-0.3cm}
\section{Evaluation}
\subsection{Experimental Setup}
\textbf{Baselines.} We compare eLLM with the following state-of-the-art LLM serving systems:
\textbf{vLLM\cite{DBLP:vllm}~}: A state-of-the-art LLM serving system widely adopted in industry.
\textbf{vLLM with Chunked Prefill\cite{DBLP:Sarathi-Serve}~}: vLLM integrates chunked prefill optimization as an optional feature, segmenting prefill requests into smaller chunks (default: 512 tokens).
\textbf{DistServe~\cite{DBLP:DistServe}}: A classical Prefill-Decode disaggregation system designed for online serving. 

\textbf{Models.} We evaluate eLLM on three popular models with distinct architectures: \textbf{Llama3-8B-262K~\cite{DBLP:llama}}: A representative Grouped Query Attention (GQA)~\cite{DBLP:GQA} model supports chunked prefill~\cite{DBLP:Sarathi-Serve}; \textbf{Jamba~\cite{DBLP:jamba}~}: A novel hybrid Mamba-MoE architecture that achieves superior performance and efficiency but does not support chunked prefill yet due to its unique design;
\textbf{OPT-13B~\cite{DBLP:opt}}: A standard Multi-Head Attention (MHA) model compatible with DistServe~\cite{DBLP:DistServe}.

\textbf{Testbed.} We evaluate eLLM and all baselines on a server configured with eight NVIDIA A100 (80GB) connected via NVLink and Intel(R) Xeon(R) Platinum 8369B CPU with 1 TB RAM.
We use PyTorch 2.4.0, CUDA Driver 12.4, and vLLM v0.5.5. 

\textbf{Workloads.} 
eLLM is evaluated under both online serving and offline inference scenarios, employing popular real-world datasets (ShareGPT~\cite{Sharegpt}) and fixed-length datasets. 
In the online serving scenario, the server is configured to issue requests according to a Poisson process with varying arrival rates. In contrast, during offline inference, all requests arrive at the beginning of the experiment~\cite{DBLP:vllm}.

\textbf{Metrics.} 
For online serving evaluation: (1) \textbf{TTFT}, Time To First Token; (2) \textbf{TPOT}, Time Per Output Token; (3) \textbf{Output Throughput}, measured in tokens/s; and (4) \textbf{SLO Attainment}, representing the percentage of requests meeting latency requirements. SLO constraints are defined as 25$\times$ the TTFT and TPOT measured under no contention~\cite{DBLP:loongserve}. From SLO Attainment results, we can also derive goodput, defined as the maximum request rate that achieves 90\% SLO attainment~\cite{DBLP:DistServe}.
For offline inference scenarios: (1) Total Throughput, measuring overall system processing capacity; (2) Decode Throughput, specifically measuring the decoding phase performance; and (3) Maximum Batch Size.

\begin{figure}[t]
    \centering
    \begin{subfigure}{0.21\textwidth}
        \centering
        \includegraphics[width=\textwidth]{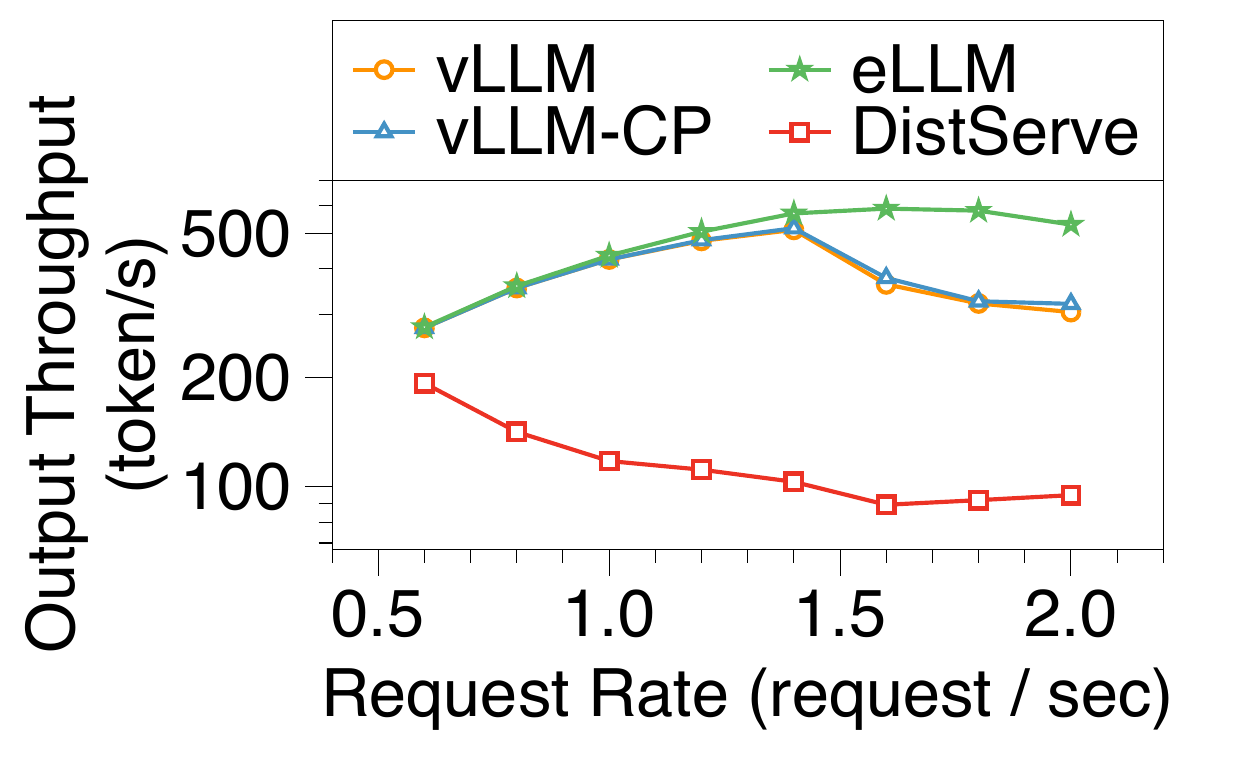}
        \captionsetup{skip=-2pt} 
        \caption{OPT-13B}
        \label{fig:optdectpslo}
    \end{subfigure}
    \hfill
    \begin{subfigure}{0.21\textwidth}
        \centering
        \includegraphics[width=\textwidth]{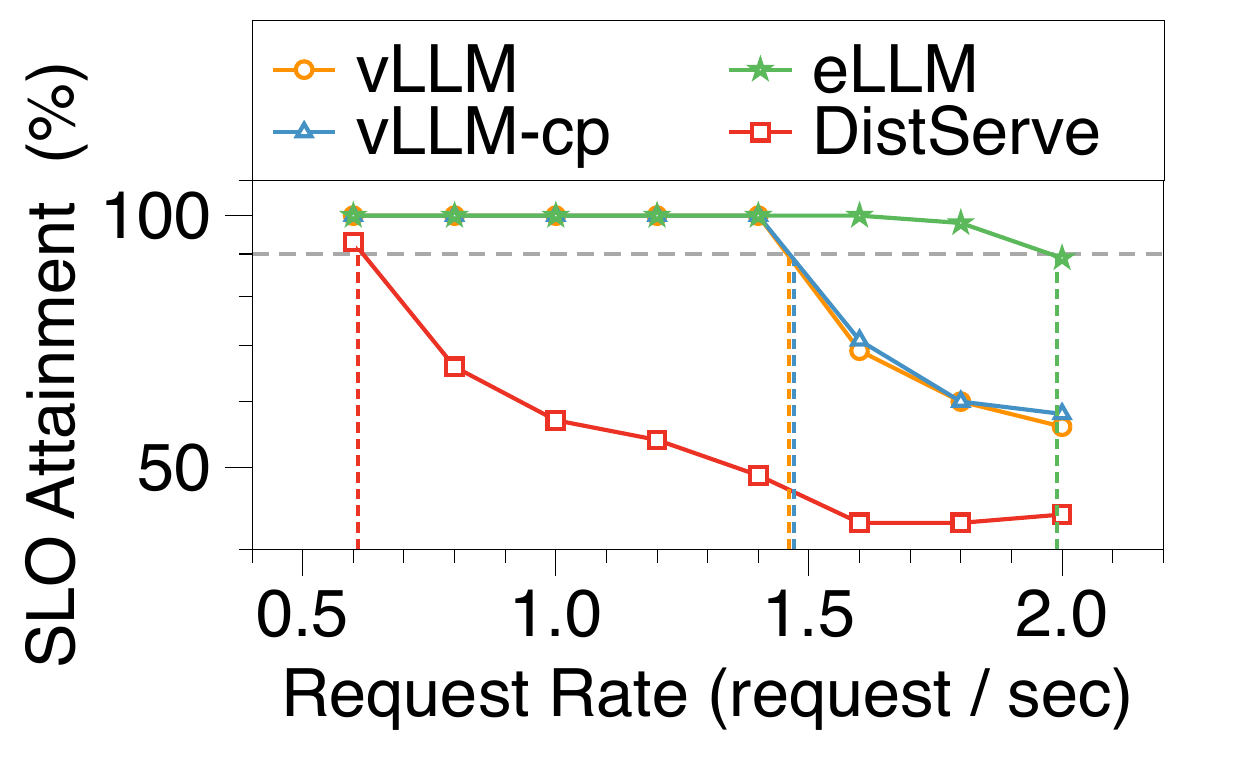}
        \captionsetup{skip=-2pt} 
        \caption{OPT-13B}
        \label{fig:optsloattain}
    \end{subfigure}
    \vspace{-0.3cm}
    \caption{SLO attainment and goodput evaluation with SLO-constraint, which is conducted on OPT-13B model with two GPUs, P=1, D=1 for DistServe and TP=2 for other systems. 
    }
    \label{fig:distserve} 
    \vspace{-0.3cm}
\end{figure}

\vspace{-0.3cm}
\subsection{Online Serving Evaluation on Single GPU}
\label{subsec:online}

Figure~\ref{fig:onlineserving} illustrates the online serving evaluation of Llama3-8B on a single A100 (80GB) GPU under various workloads. Figure~\ref{fig:2k2kttft}, and Figure~\ref{fig:shttft} showcase the TTFT performance of eLLM under different workloads.
eLLM consistently outperforms vLLM across both workloads, with up to 295$\times$ and 140$\times$ faster TTFT compared to vLLM and vLLM-CP respectively.

Specifically, in Figure~\ref{fig:2k2kttft}, vLLM-CP and eLLM both exhibit its ability to reduce queueing time. However, with the request rate increasing, vLLM-CP's performance degrades. In contrast, eLLM makes full use of the intra-GPU elasticity and leverages the GPU-CPU elasticity to reduce the queueing time, leading to the best performance.
The larger decoding batch of eLLM also contributes to the improved output, as shown in Figure~\ref{fig:shout}, and Figure~\ref{fig:2k2kout}.

For TPOT metric, because eLLM and vLLM-CP have more available KV Cache space for decoding, their TPOT is higher than vLLM.
Overall, the trade-off of eLLM between TTFT and TPOT leads to a better SLO attainment and up to 2.5$\times$ and 2.26$\times$ higher goodput compared to vLLM and vLLM-CP in Figure~\ref{fig:2k2kslo} and Figure~\ref{fig:shslo}.

\vspace{-0.3cm}
\subsection{Online Serving Evaluation on Multi-GPUs}

We evaluated the online performance of eLLM in multi-GPU scenarios, as shown in Figure~\ref{fig:distserve}, allowing us to include DistServe with prefill-decoding disaggregation for comparison. 
eLLM consistently achieves the highest performance across all tested configurations. Although DistServe theoretically allows decoding instances to minimize pre-allocated activation space, its actual performance remains suboptimal due to two main factors:
First, DistServe executes prefill and decoding phases separately on dedicated GPUs. Consequently, computational resources on one GPU are underutilized when the corresponding phase is idle.
Second, model weights in DistServe are replicated across GPUs, resulting in additional memory consumption and a reduction in the effective batch size during decoding.

\vspace{-0.3cm}
\subsection{Offline Inference Evaluation}

\begin{figure}[t]
    \centering
    \includegraphics[width=0.30\textwidth]{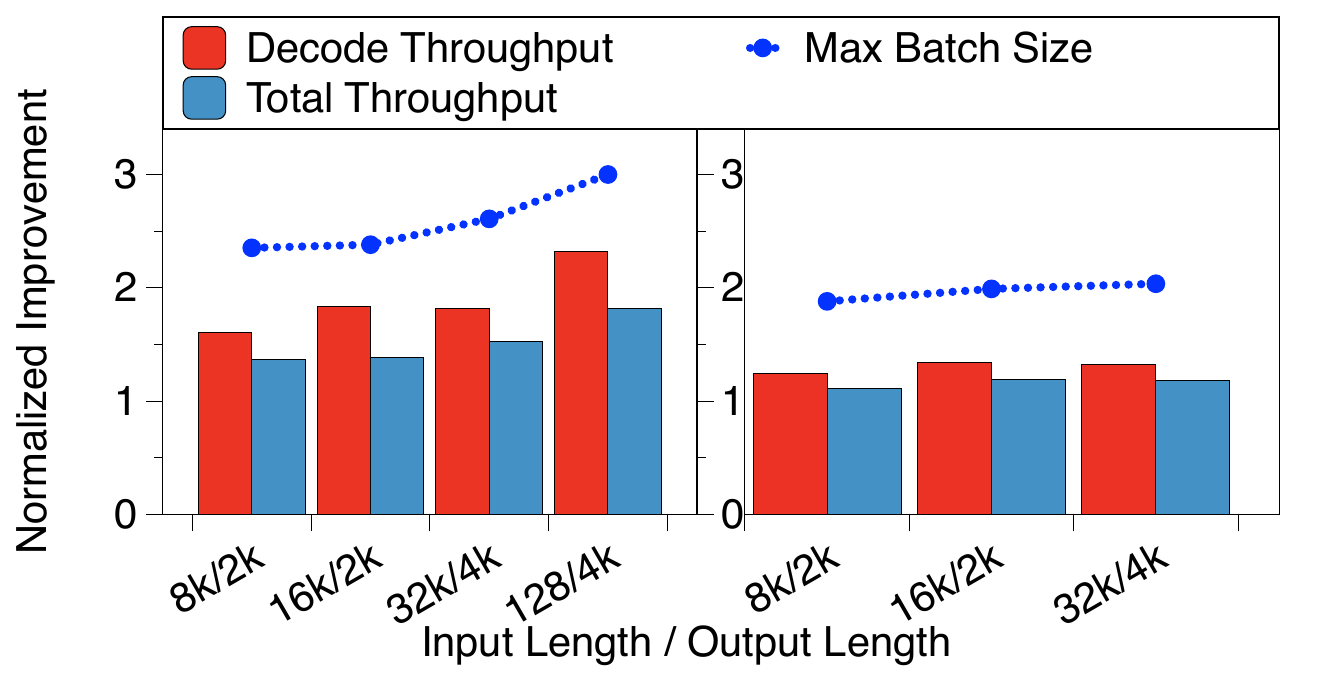} 
    \captionsetup{skip=-2pt} 
    \caption{Throughput and batch size performance normalized to vLLM, under varying input/output lengths. Left: Jamba-Mini~\cite{DBLP:jamba} on 2 GPUs; Right: Llama3-8B-262K on 1 GPU.}
    \label{fig:offline_test} 
\end{figure}

Figure~\ref{fig:offline_test} shows eLLM's offline inference performance across various configurations. By dynamically converting idle activation memory to KV cache during decoding, eLLM maintains larger batch sizes than vLLM. The benefits are particularly pronounced for long sequences and architectures with high activation-to-KV ratios like Jamba, where vLLM's static allocation causes severe memory bottlenecks. eLLM achieves up to 1.82$\times$ total throughput and 2.32$\times$ decode throughput improvements in the 128k-4k configuration.

\vspace{-0.3cm}
\subsection{Ablation Study}

\begin{figure}[t]
    \centering
    \begin{subfigure}{0.21\textwidth}
        \centering
        \includegraphics[width=\textwidth]{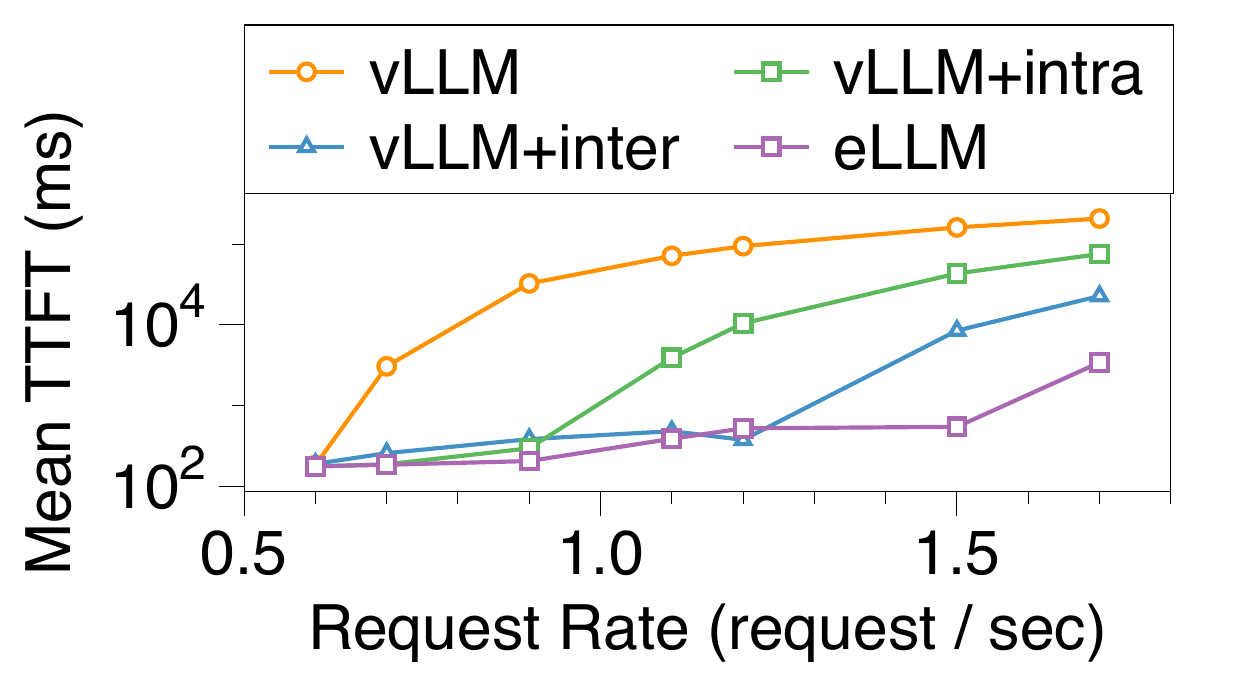}
        \captionsetup{skip=-2pt} 
        \caption{TTFT}
        \label{fig:a}
    \end{subfigure}
    \hfill
    \begin{subfigure}{0.21\textwidth}
        \centering
        \includegraphics[width=\textwidth]{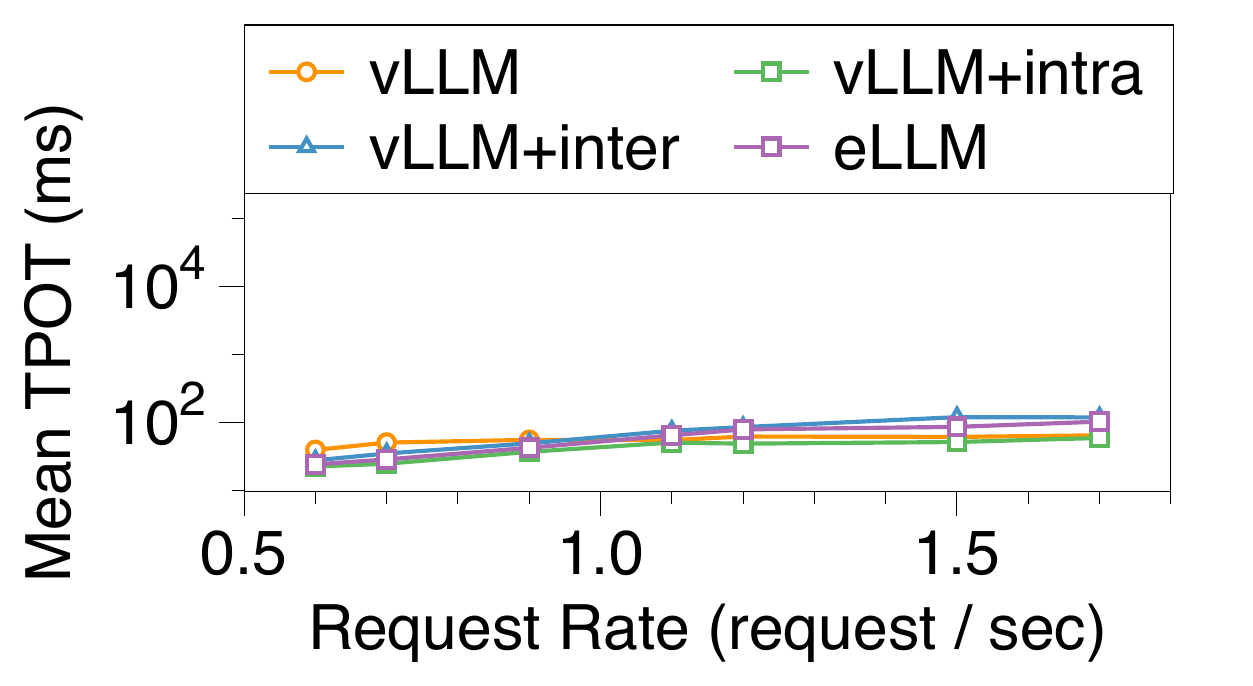}
        \captionsetup{skip=-2pt} 
        \caption{TPOT}
        \label{fig:b}
    \end{subfigure}
    \hfill
    \begin{subfigure}{0.21\textwidth}
        \centering
        \includegraphics[width=\textwidth]{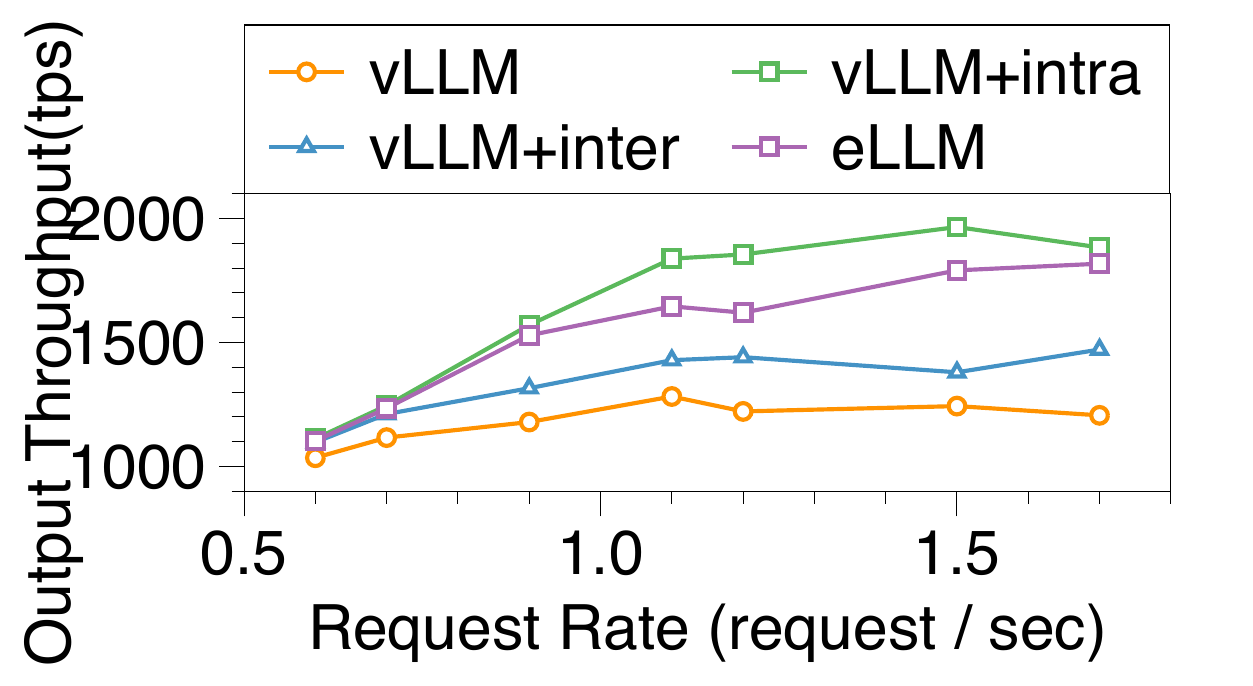}
        \captionsetup{skip=-2pt} 
        \caption{Output Throughput}
        \label{fig:c}
    \end{subfigure}
    \hfill
    \begin{subfigure}{0.21\textwidth}
        \centering
        \includegraphics[width=\textwidth]{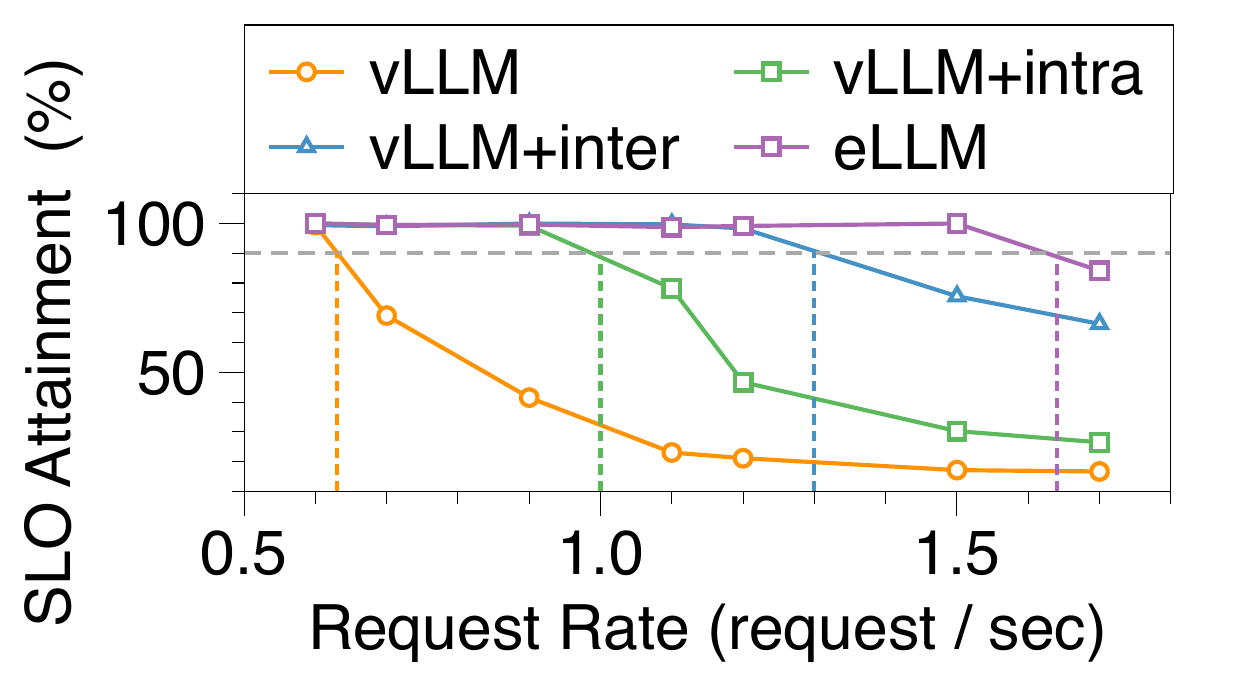}
        \captionsetup{skip=-2pt} 
        \caption{SLO Attainment}
        \label{fig:d}
    \end{subfigure}
    \vspace{-0.3cm}
    \caption{Ablation study on different features of eLLM. This experiment is conducted on the 2K-2K workload with the same settings as the online serving evaluation.}
    \label{fig:ablation} 
    \vspace{-0.3cm}
\end{figure}

We conducted an ablation study on the two elastic features of eLLM: intra-GPU elasticity (denoted as \texttt{vLLM+intra}) and GPU-CPU elasticity (denoted as \texttt{vLLM+inter}). Using vLLM as the baseline, we evaluated each method's effectiveness, as shown in Figure~\ref{fig:ablation}.

Figure~\ref{fig:a} demonstrates that both intra-GPU and GPU-CPU elasticity mechanisms reduce TTFT by expanding effective memory capacity for incoming requests. While each mechanism provides substantial improvements independently, their combination in eLLM achieves up to 295$\times$ TTFT reduction through synergistic memory management. Importantly, these gains are achieved without degrading TPOT stability (Figure~\ref{fig:b}).
Figure~\ref{fig:c} and~\ref{fig:d} reveal critical performance trade-offs. Although both mechanisms increase decoding batch sizes, their combination sometimes yields lower throughput than using intra-GPU elasticity alone, primarily due to PCIe transfer overhead that cannot be fully overlapped in decoding phase. Moreover, intra-GPU elasticity alone fails to prevent prefill-phase bottlenecks under high load, resulting in goodput degradation. eLLM addresses these limitations through dynamic dual-mechanism orchestration, achieving 1.5$\times$ throughput and 2.5$\times$ goodput improvements by adaptively balancing the benefits of GPU-internal memory sharing against the costs of CPU-backed overflow.

\vspace{-0.3cm}
\subsection{System Execution Time Breakdown}

We quantify eLLM's overhead by decomposing execution time into three components: CPU scheduling, VMM operations, and model execution. 
The CPU scheduling overhead remains negligible (less than 1\% of total execution time) due to our lightweight scheduling strategy. 
VMM operation overhead, comprising eLLM’s internal calls to VMM API functions, accounts for 1-5\% of execution time in online serving scenarios. This modest overhead is achieved through three optimization techniques: eTensor pooling to reuse memory mappings, speculative pre-mapping for decoding phases, and asynchronous unmapping to overlap memory operations with computation. These optimizations ensure that eLLM's memory virtualization benefits far outweigh its runtime costs.

\vspace{-0.3cm}
\section{Related Work}
Existing research on long-context LLM memory challenges generally falls into three categories:
\textbf{Lightweight Algorithms.} Prior work reduces memory demands by optimizing attention architecture (e.g., shared head attention~\cite{DBLP:GQA}, hybrid transformer~\cite{DBLP:jamba}) and compression techniques (e.g., quantization~\cite{DBLP:KIVI,DBLP:KVQuant}, sparsification~\cite{DBLP:H2O,DBLP:StreamingLLM,DBLP:Keyformer}).
Although effective in reducing memory usage, these approaches often compromise model accuracy.
\textbf{Efficient Kernels.} Existing LLM serving systems utilize optimized GPU kernels that take advantage of memory hierarchies to improve memory access efficiency and reduce memory footprints. Examples include Flash Attention~\cite{DBLP:FA} and Flash-Decoding~\cite{DBLP:FlashDecoding}. These techniques are orthogonal to our approach and have been integrated into eLLM.
\textbf{Memory Management.} GMLake~\cite{DBLP:GMLake} addresses fragmentation issues caused by the BFC algorithm in memory pools, while vAttention~\cite{prabhu2025vattentiondynamicmemorymanagement} tackles fine-grained fragmentation at the token dimension. 
While these approaches effectively reduce memory fragmentation within their respective scopes, they maintain the fundamental isolation between activation and KV cache memory spaces, which limits resource utilization and hinders dynamic workload management. 
In contrast, eLLM fundamentally addresses space-wise internal fragmentation by enabling elastic memory sharing between previously isolated spaces, representing a paradigm shift in memory management for long-context LLM serving.

\vspace{-0.3cm}
\section{Conclusion}
Modern LLM serving systems face significant memory utilization bottlenecks due to the isolation of memory spaces.
Our proposed \texttt{eLLM} framework addresses this through an elastic memory paradigm. By introducing virtual tensor abstraction to unify memory pools, dynamic inflation/deflation mechanisms to optimize GPU utilization, and CPU-GPU memory orchestration with SLO-aware scheduling, \texttt{eLLM} achieves significantly improvements. 
\vspace{-0.3cm}
\begin{acks}
This work was supported by Fundamental and Interdisciplinary Disciplines 
Breakthrough Plan of the Ministry of Education of China (JYB2025XDXM113) 
and National Natural Science Foundation of China (NSFC) grant (62532006).
\end{acks}


\bibliographystyle{ACM-Reference-Format}
\bibliography{acmart}

\end{document}